\documentclass[sigconf]{acmart}
\AtBeginDocument{%
  }

\setcopyright{acmlicensed}
\copyrightyear{2018}
\acmYear{2018}
\acmDOI{XXXXXXX.XXXXXXX}
\acmConference[Conference acronym 'XX]{Make sure to enter the correct
  conference title from your rights confirmation email}{June 03--05,
  2018}{Woodstock, NY}
\acmISBN{978-1-4503-XXXX-X/2018/06}

\usepackage{multirow}
\usepackage{graphicx}
\usepackage{tabularx}
\usepackage{mathtools}

\begin{document}
\title{Retrieval-GRPO: A Multi-Objective Reinforcement Learning Framework for Dense Retrieval in Taobao Search}

\author{Xingxian Liu}
\email{liuxingxian.lxx@alibaba-inc.com}
\affiliation{%
  \institution{Taobao \& Tmall Group of Alibaba}
  \city{Hangzhou}
  \country{China}
}

\author{Dongshuai Li}
\email{lidongshuai.lds@alibaba-inc.com}
\affiliation{%
  \institution{Taobao \& Tmall Group of Alibaba}
  \city{Hangzhou}
  \country{China}
}

\author{Jiahui Wan}
\email{wanjiahui.wjh@alibaba-inc.com}
\affiliation{%
  \institution{Taobao \& Tmall Group of Alibaba}
  \city{Hangzhou}
  \country{China}
}

\author{Tao Wen}
\email{wentao.wen@alibaba-inc.com}
\affiliation{%
  \institution{Taobao \& Tmall Group of Alibaba}
  \city{Hangzhou}
  \country{China}
}

\author{Gui Ling}
\email{linggui.lg@alibaba-inc.com}
\affiliation{%
  \institution{Taobao \& Tmall Group of Alibaba}
  \city{Hangzhou}
  \country{China}
}

\author{Yuliang Yan}
\email{yuliang.yyl@alibaba-inc.com}
\affiliation{%
  \institution{Taobao \& Tmall Group of Alibaba}
  \city{Hangzhou}
  \country{China}
}

\author{Fuyu Lv}
\authornote{Corresponding author.}
\email{fuyu.lfy@alibaba-inc.com}
\affiliation{%
  \institution{Taobao \& Tmall Group of Alibaba}
  \city{Hangzhou}
  \country{China}
}

\author{Dan Ou}
\email{oudan.od@alibaba-inc.com}
\affiliation{%
  \institution{Taobao \& Tmall Group of Alibaba}
  \city{Hangzhou}
  \country{China}
}

\author{Haihong Tang}
\email{piaoxue@taobao.com}
\affiliation{%
  \institution{Taobao \& Tmall Group of Alibaba}
  \city{Hangzhou}
  \country{China}
}

\author{Bo Zheng}
\email{bozheng@alibaba-inc.com}
\affiliation{%
  \institution{Taobao \& Tmall Group of Alibaba}
  \city{Beijing}
  \country{China}
}


\begin{abstract}
Dense retrieval, as the core component of e-commerce search engines, maps user queries and items into a unified semantic space through pre-trained embedding models to enable large-scale, real-time semantic retrieval. Despite the rapid advancement of large language models (LLMs) gradually replacing traditional BERT architectures for embedding, their training paradigms still adhere to BERT-like supervised fine-tuning and hard negative mining strategies. This approach relies on complex offline hard negative sample construction pipelines, which constrain model iteration efficiency and hinder the evolutionary potential of semantic representation capabilities. Furthermore, existing multi-task learning frameworks face the seesaw effect when simultaneously optimizing semantic relevance and non-relevance objectives, limiting model performance ceilings. In this paper, we propose Retrieval-GRPO, a multi-objective reinforcement learning-based dense retrieval framework designed to address these challenges. The method eliminates offline hard negative sample construction by dynamically retrieving Top-K candidate products for each query during training, while introducing a relevance LLM as a reward model to generate real-time feedback. Specifically, the retrieval model dynamically optimizes embedding representations through reinforcement learning, with reward signals combining LLM-generated relevance scores, product quality scores, and multi-way exclusivity metrics to achieve multi-objective user preference alignment and real-time error correction. This mechanism not only removes dependency on hard negatives but also mitigates the seesaw effect through collaborative multi-objective optimization, significantly enhancing the model's semantic generalization capability for complex long-tail queries. Extensive offline and online experiments validate the effectiveness of Retrieval-GRPO, which has been deployed on China's largest e-commerce platform.
\end{abstract}

\ccsdesc[500]{Information systems~Language models}

\keywords{Dense Retrieval, Large Language Model, Reinforcement Learning, E-commerce Search System}



\maketitle

\section{Introduction}
Dense retrieval\cite{Guo_Fan_Ai_Croft_2016}\cite{Guo_Fan_Pang_Yang_Ai_Zamani_Wu_Croft_Cheng_2020}\cite{Mitra_Craswell_2017} has emerged as a critical technology in e-commerce search engines to address the unique challenge of precisely aligning massive items with user queries. This approach employs advanced pre-trained embedding models to encode user queries and item descriptions into a unified semantic vector space, enabling real-time semantic search over billions of items via Approximate Nearest Neighbor (ANN) algorithms\cite{malkov2018efficient}. The methodology effectively learns query-item alignment relationships within a unified semantic space, even in the absence of direct lexical overlap between them. Dense retrieval directly improves users' search experience and merchant exposure efficiency, thereby driving significant improvements in overall platform transaction conversion rates.

Previous approaches primarily relied on encoder-only architectures like BERT\cite{Devlin_Chang_Lee_Toutanova_2019} and RoBERTa\cite{Liu_Ott_Goyal_Du_Joshi_Chen_Levy_Lewis_Zettlemoyer_Stoyanov_2019}, utilizing the bidirectional attention mechanism to produce contextual semantic embeddings. Benefiting from rapid advancements in large language model technologies, emerging approaches such as Gemini Embedding\cite{lee2025gemini}, Seed Embedding\cite{guo2025seed1}, QWEN3-Embedding\cite{zhang2025qwen3}, and NV-Embed\cite{lee2024nv} employ LLMs as encoders to generate semantic embeddings for retrieval, leveraging their extensive knowledge bases and zero-shot generalization capabilities. Despite incorporating LLMs, most current methods still adopt training paradigms designed for BERT-like models, enhancing performance on long-tail challenging queries through continued supervised fine-tuning (SFT) and persistent hard negative sample mining. Notably, offline hard negative sample mining constitutes a complex systemic workflow, which restricts iteration speed and limits the evolution potential of representation capabilities.

Furthermore, in e-commerce search systems, semantic relevance between queries and items is not the sole evaluation criterion. Existing methods typically employ multi-task learning frameworks when modeling non-relevance objectives. However, the simultaneous optimization of multiple losses suffers from the seesaw phenomenon\cite{10.1145/3383313.3412236}, which constrains the model's performance ceiling.

In this paper, we propose Retrieval-GRPO, a multi-objective reinforcement learning framework for dense retrieval. This approach eliminates the need for offline manual construction of hard negative samples. Instead, during training, we dynamically retrieve top-k items for each query and employ a relevance LLM\cite{dong2025taosr1} as the reward model to generate scores. These relevance scores are combined with item quality scores and exclusivity metrics to form multi-objective rewards, enabling real-time user preference alignment and error correction through RL training.

In summary, compared to conventional dense retrieval paradigms, Retrieval-GRPO offers three key contributions:

\begin{itemize}
\item Eliminates laborious offline hard negative sample mining, resolving protracted iteration cycles between sample selection and training.
\item Utilizes candidates sampled by the latest gradient-updating model during training, whereas offline hard negatives derive from static older models. This enables instant learning from relevance model feedback for real-time error correction.
\item Multi-objective reward fusion provides more direct and flexible optimization compared to multi-loss approaches, effectively mitigating the seesaw effect in multi-task training.
\end{itemize}

\section{Related Works}
\subsection{Dense Retrieval}
Dense retrieval encodes queries and items into a shared semantic space and retrieves the top-N most similar items using approximate nearest neighbor (ANN) algorithms\cite{malkov2018efficient}. Early dense retrieval methods primarily relied on encoder-only pre-trained models like BERT\cite{Devlin_Chang_Lee_Toutanova_2019}, RoBERTa\cite{Liu_Ott_Goyal_Du_Joshi_Chen_Levy_Lewis_Zettlemoyer_Stoyanov_2019}. SimCSE\cite{gao2021simcse} first introduced a contrastive learning framework for dense retrieval tasks, using dropout as an unsupervised self-prediction technique in its contrastive 
learning framework. After that, 
the contrastive learning paradigm became widely adopted. Beyond pure text representation, CLIP\cite{radford2021learning} leverages a large-scale image-text 
matching dataset to pretrain multimodal representations through simple contrastive learning objectives for multimodal alignment tasks. With the rapid development of large language models, many studies now employ LLMs as backbones for dense retrieval, leveraging their extensive world knowledge and zero-shot generalization capabilities to obtain superior universal semantic representations. Notably, RepLLaMA\cite{ma2024fine} first demonstrated the effectiveness of multi-stage supervised fine-tuning (SFT) of large language models for text retrieval. NV-Embed\cite{lee2024nv} proposes a latent attention layer to obtain pooled embeddings and enhances model performance through contrastive learning with in-batch negatives and curated hard negative examples. Gemini Embedding\cite{lee2025gemini} utilizes Gemini's multilingual and code comprehension capabilities to generate highly versatile embeddings across languages and text modalities. Seed Embedding\cite{guo2025seed1} adopts a Mixture-of-Experts (MoE) model as its backbone, supporting multiple embedding dimensions via Matryoshka Representation Learning (MRL)\cite{kusupati2022matryoshka} training similar to NV-Embed, and employs two-stage contrastive learning to strengthen universal representation capabilities. The gte-Qwen\cite{li2023towards} integrates bidirectional attention mechanisms and enhances performance through instruction tuning using both weakly supervised and supervised data. Building on gte-Qwen, Qwen3 Embedding\cite{zhang2025qwen3} introduces an innovative multi-stage training pipeline that combines large-scale unsupervised pre-training with supervised fine-tuning on high-quality datasets, ensuring robustness and adaptability through effective model merging strategies. Current LLM-based dense retrieval models achieve state-of-the-art (SOTA) performance on the MTEB leaderboard\cite{muennighoff2023mteb}, demonstrating the effectiveness of large models for semantic representation.

\begin{figure*}[t]
    \centering
    \includegraphics[width=\linewidth]{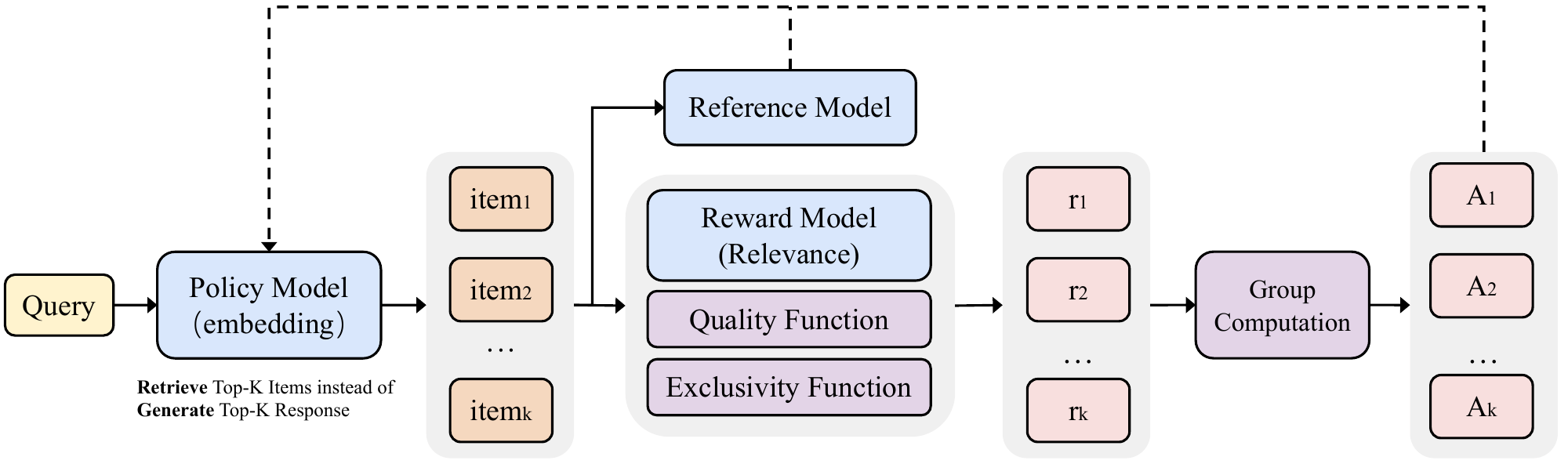}
    \caption{The Retrieval-GRPO overall training framework consists of 3 steps: (1)Candidates Selection; (2)Multi-Objective Reward Calculation; (3)GRPO Optimization}
    \label{fig:pipeline} 
\label{pic:main}
\end{figure*}

\subsection{Reinforcement Learning for LLM}
In recent years, large language model (LLM) technology has rapidly advanced, with reinforcement learning (RL)\cite{wang2024reinforcement} being widely integrated into LLM training pipelines as a key technical approach to enhance model performance and controllability. The core application of RL in LLMs involves modeling text generation as a sequential decision-making problem, where the model acts as an agent, generated text represents actions, and external feedback forms reward signals. Among these approaches, Reinforcement Learning from Human Feedback (RLHF) has quickly become the dominant paradigm. OpenAI's Proximal Policy Optimization (PPO)\cite{Schulman_Wolski_Dhariwal_Radford_Klimov_2017} stands as one of the most representative algorithms in the RLHF framework. Subsequently, OpenAI first applied PPO to align human preferences in its InstructGPT\cite{ouyang2022training}. In PPO, the policy model serves as the optimization target, generating sampled text from input prompts. The reward model, trained on human preference data, provides scalar reward signals based on response quality from the policy model. A reference model calculates KL divergence penalties to prevent excessive deviation from the original model, while a value model estimates action value functions representing expected cumulative future value. The PPO algorithm coordinates these four components to align human preferences. Despite PPO's success, its complex workflow and high computational costs persist. The Direct Preference Optimization (DPO)\cite{rafailov2023direct} algorithm addresses this by assuming human preferences follow the Bradley-Terry model, simplifying preference learning into a directly optimizable classification problem through maximum likelihood objectives. While DPO achieves stable performance with higher efficiency and outperforms PPO in some tasks, PPO maintains advantages in complex multi-step reasoning tasks. To reduce PPO's complexity and computational demands, DeepSeek proposes the GRPO\cite{shao2024deepseekmath} algorithm during the training of its DeepSeekMath. GRPO eliminates the value model in PPO and adopts a group-sampling-based baseline estimation method, using average group rewards as the baseline to calculate inter-group advantages, significantly reducing computational overhead. Building on PPO, LitePPO\cite{liu2025part} discovered that simply applying robust advantage normalization and token-level loss aggregation techniques to vanilla PPO could consistently improve performance. Currently, the RL research continues to drive progress in LLM training, providing sustainable iterative methods for aligning large language models with human preferences.

\section{Methods}
\label{sec:preliminary}
In modern e-commerce search systems, dense retrieval aims to map queries and items into a shared semantic embedding space, enabling efficient selection through geometric proximity. Given a query \( q_i \) and an item collection \( \mathcal{D} = \{d_1, \ldots, d_n\} \), the framework typically employs a dual-encoder architecture\cite{reimers2019sentence} to encode \( q_i \) and \(\mathcal{D}\) to a semantic embedding as Equation \ref{eq:score1} and \ref{eq:score2}.
\begin{equation}
\mathbf{q}_i = f_\theta(q_i)
\label{eq:score1}
\end{equation}
\begin{equation}
\mathbf{d}_j = f_\theta(d_j)
\label{eq:score2}
\end{equation}
where \( f_\theta(\cdot) \) denotes a pretrained model that independently projects text into fixed-dimensional vectors. The retrieval process computes semantic similarity scores \( s(\mathbf{q}_i, \mathbf{d}_j) \in \mathbb{R} \) through measures(e.g., cosine similarity) between \(\mathbf{q}_i\) and \(\mathbf{d}_j\).

Relevant items \( \mathcal{D}^+_{q_i} = \{d^+_{q_i,1}, \ldots, d^+_{q_i,m}\} \) (\( m \ll n \)) are retrieved via:
\begin{equation}
    \mathcal{D}_{q_i} = \mathop{\mathrm{TopK}}_{d_j \in \mathcal{D}} \, s(\mathbf{q}_i, \mathbf{d}_j)
\end{equation}

In this section, we detail our multi-objective dense retrieval training framework, which includes Supervised Fine-Tuning (SFT) and Retrieval-GRPO. Additionally, we provide a 
detailed, separate
introduction to the Reward design in Retrieval-GRPO. First, SFT is performed to endow the model with discriminative capabilities on simple positive and negative samples. Then, through Retrieval-GRPO, the model learns from the reward of multi-objective user preferences for real-time correction, eliminating the need for traditional manual construction of hard negative samples\cite{li2024syneg}.

\subsection{SFT}
\label{method:sft}
Traditionally, in the SFT phase of dense retrieval, training samples consist of <query, item> pairs for positive samples, while negative samples are not specifically collected but obtained through in-batch negative sampling\cite{radford2021learning}. The InfoNCE loss is used to pull closer the distance between positive query-item pairs and push away irrelevant query-item pairs in the semantic embedding space. The InfoNCE loss is 
as follows:
\begin{equation}
\mathcal{L}_{\text{InfoNCE}} = -\log \frac{\exp(s(q_i, d_j^+) / \tau)}{ \sum_{d_j \in \mathcal{B}} \exp(s(q_i, d_j) / \tau)}
\end{equation}
where \(s(\cdot)\) denotes similarity scoring, \(\tau\) is the temperature parameter, and \(\mathcal{B}\) is the in-batch negatives, 

However, this approach leads to only those items present as positive samples in the training set being utilized as negative samples for other queries. Meanwhile, items in the product pool that lack historical user interactions are entirely ignored during training. To address this, we enhance the conventional InfoNCE in-batch negative sampling by introducing global negative sampling, which randomly selects items from the product pool to serve as negative samples for queries. Consequently, the negative samples are derived partially from in-batch sampling and partially from the product pool, ensuring that all items in the product pool are exposed to the model during training.
The loss is 
as follows:
\begin{equation}
\mathcal{L} = -\log \frac{\exp(s(q_i, d_j^+) / \tau)}
{ 
    \underbrace{\textstyle\sum_{d_j \in \mathcal{B}} \exp(s(q_i, d_j) / \tau)}_{\mathclap{\text{\scriptsize positive and in-batch negatives}}} + 
    \underbrace{\textstyle\sum_{d_k^- \in \mathcal{G}} \exp(s(q_i, d_k^-) / \tau)}_{\mathclap{\text{\scriptsize global negatives}}}
}
\end{equation}
where \(\mathcal{G}\) represents global negatives randomly sampled from the global item-pool. This dual-sampling strategy ensures all items in the pool participate in training, with negatives comprising both in-batch \(\mathcal{B}\) and global \(\mathcal{G}\) query-item pairs.

\subsection{Retrieval-GRPO}
After the SFT phase, the dense retrieval model should acquire the capability to distinguish between simple positive and negative samples. In the previous training framework, hard negative samples are typically introduced to enhance model performance on long-tail challenging queries. 

However, this requires offline manual construction of hard negative samples, and the hard negative mining process itself constitutes a complex systemic task, which significantly hinders further scalability of model capabilities.
To address this, we propose Retrieval-GRPO, a reinforcement learning paradigm specifically designed for dense retrieval, which eliminates the need for manual hard-negative samples mining.

The overall training pipeline of Retrieval-GRPO is illustrated in Figure \ref{pic:main} and consists of three steps:

\begin{itemize}
\item \textbf{Candidate Selection}: The dense retrieval model encodes queries and items into a unified semantic embedding space. Using the Approximate Nearest Neighbor (ANN) algorithm\cite{malkov2018efficient}, it retrieves the top-k items with the closest distances as candidate items.
\item \textbf{Reward Computation \(\&\) Fusion}: Rewards are defined based on relevance, item quality, and exclusivity. These reward scores are then integrated into a multi-objective reward that combines perspectives from user preferences, product supply, and the search system.
\item \textbf{GRPO Loss Optimization}: The inter-group advantage among different item groups under the same query is calculated. The vector retrieval model is optimized using the GRPO loss, as defined in Equation \ref{eq:grpo}.
\end{itemize}

\subsubsection{Multi-Objective Reward}
\label{method_mul_reward}
Within the reinforcement learning paradigm, the most critical question to address is how to design the reward mechanism, specifically determining which retrieved results should be rewarded or penalized.
We answer this question through three dimensions: relevance, quality, and exclusivity. Relevance between queries and items reflects the user experience perspective, item quality represents the objective supply perspective of products, and exclusivity arises from the system-level considerations of multi-way retrieval.

To satisfy users' search experience, dense retrieval 
aims 
to prioritize items that are highly relevant to the user's query and of superior quality. However, due to the prevalence of multi-way retrieval architectures in e-commerce search systems, we particularly emphasize incrementally exclusive relevant items—those that are semantically related to the query but not exact matches—that provide incremental value compared to results from other retrieval ways.

Thus, the multi-objective reward in our reinforcement learning process is jointly defined by relevance, quality, and exclusivity:
\begin{equation}
r = f_{\text{relevance}}(q, d) + g_{\text{quality}}(d) + h_{\text{exclusivity}}(q, d)
\end{equation}
where the \(f_{\text{relevance}}(q, d)\) denotes the relevance score between query and item computed by TaoSR1, \(g_{\text{quality}}(d)\) denotes
the objective quality score of items derived from historical transaction data, and 
\(h_{\text{exclusivity}}(q, d)\) measures exclusivity against other retrieval-ways, indicating whether the item provides incremental value by being retrieved through the current channel but not through alternative retrieval ways.

\begin{itemize}
\item \textbf{Relevance reward}:The query-item relevance score is computed by TaoSR1\cite{dong2025taosr1}, an advanced internal downstream relevance model deployed independently. TaoSR1 is a 42B-MoE LLM trained through multi-stage RL learning. We leverage it as the reward model in our reinforcement learning framework to provide real-time relevance scoring for query-item candidates generated by dense retrieval model.
\item \textbf{Quality reward}: A discrete value representing the objective item quality score, derived from historical transaction data and user satisfaction metrics.
\item \textbf{Exclusivity reward}: Measures the overlap with other retrieval ways. Initially a posterior metric, this is simplified to a prior metric due to the dominance of inverted-index-based retrieval in the e-commerce search system: If an item is retrieved by the inverted-index channel (i.e., has a high term overlap ratio between query and item), it receives a lower exclusivity reward.
\end{itemize}

\subsubsection{Candidates Selection}

In the Retrieval-GRPO framework, instead of offline hard negative samples construction, we treat the top-k items retrieved by the dense retrieval model as candidate predictions requiring correction. Leveraging the relevance reward model and multi-objective reward function described in Section \ref{method_mul_reward}, the rewards 
are computed in real time. Items within the top-k that receive low reward scores are penalized to dynamically adjust the model's predictions.

However, identifying top-k items for every query across the entire product pool is computationally intensive. We simplify this process by gathering items across devices to form a large batch \(\hat{B}\) (where \(k\ll\hat{B}\)), then selecting the top-k candidates from this batch for reward computation.

\subsubsection{GRPO loss}
As the reward computation described in Section \ref{method_mul_reward}, we perform multi-objective reward calculation using the reward model and reward function on the top-k items retrieved by the dense retrieval model. These rewards are fused into a unified multi-objective reward that integrates perspectives from user preferences, item supply, and the search system.

Following the GRPO training paradigm, we compute the inter-group advantage \(A_{ij}\) for the top-k items under the same query using their rewards.\(A_{ij}\) reflects multi-perspective user preferences, indicating which top-ranked items should be rewarded or penalized for a given query. The penalized samples can be viewed as hard negative samples in the traditional sense, but the penalty is assigned as a continuous value rather than a hard label. The dense retrieval model learns from this real-time multi-objective feedback to improve its ability to distinguish longtail corner cases. The final loss is defined in Equation \ref{eq:grpo}:

\begin{equation}
\scalebox{0.9}{$
\begin{split}
    \mathcal{J}_{GRPO}(\theta) 
    &= \mathbb{E}
    \Bigg[\frac{1}{G} \sum_{i=1}^G
    \bigg\{
    \min \bigg[
    \frac{\pi_\theta(s(q,d_{i})|q, d_{i})}{\pi_{\theta_{\text{old}}}(s(q,d_{i})|q, d_{i})} \hat{A}_{i,t}, \\
    &\quad\quad \text{clip} \left( 
    \frac{\pi_\theta(s(q,d_{i})|q, d_{i})}{\pi_{\theta_{\text{old}}}(s(q,d_{i})|q, d_{i})}, 1 - \epsilon, 1 + \epsilon 
    \right) \hat{A}_{i,t} 
    \bigg] 
    - \beta \mathbb{D}_\text{KL}\big[\pi_\theta || \pi_\text{ref}\big]
    \bigg\} 
    \Bigg]
\end{split}
$}
\label{eq:grpo}
\end{equation}

where \(G\) is the size of group, \(s(q,d_{i})\) is the similarity scores calculated by policy model. \(\pi_{\theta}(\cdot)\) is the distribution of policy model, \(\pi_{\theta_{old}}(\cdot)\) is the distribution of the model which is used for sampling, \(\pi_{ref}(\cdot)\) is the distribution of reference model.

\begin{table*}[t]
    \centering
    \caption{Offline Performance Comparison of Training Strategies with Goodrate@100}
    \label{tab:main_result}
    \begin{tabularx}{0.8\textwidth}{l *{5}{>{\centering\arraybackslash}X}}
    \toprule
        \multirow{2}{*}{Model} & \multicolumn{5}{c}{Goodrate@100} \\
        & Q\&A & Alternative & Negative & Knowledge & Overall\\ 
        \midrule
        BERT-base & 69.60 & 26.86 & 49.90 & 50.49 & 49.21 \\
        Query-Rewrite & 84.52 & 36.39 & 57.40 & 62.65 & 60.24 \\
        Qwen-3B & 82.03 & 34.65 & 59.87 & 58.09 & 58.66 \\ 
        Tbstars Post-Pretrain(base) & 80.67 & 32.87 & 60.26 & 63.20 & 59.25 \\ 
        \midrule
        Tbstars SFT Only & 84.01 & 38.70 & 62.60 & 68.45 & 63.44 \\ 
        Tbstars SFT + Ranking loss(Hard Neg) & 84.90 & 40.12 & 62.40 & 68.71 & 64.03 \\ 
        Tbstars SFT + DPO(Hard Neg) & 84.96 & 40.21 & 63.60 & 68.76 & 64.38 \\ 
        Tbstars SFT + R-GRPO & 85.05 & 43.32 & 65.80 & 69.50 & 65.91 \\ 
    \bottomrule
    \end{tabularx}
\label{table:main_goodrate}
\end{table*}

\section{Experiment}
\subsection{Experiment Setup}
\label{exp.setup}
\subsubsection{Dataset and Metrics}
For offline evaluation, we construct three test sets: 
the general retrieval test set, the long-tail query evaluation set, the and long-tail experience evaluation set,
all collected and filtered from Taobao’s online search logs. The model is evaluated using Hitrate@6k and Goodrate@100.

\begin{itemize}
\item \textbf{Hitrate@6k}: A retrieval metric calculating the proportion of ground-truth items covered by the top-6k results retrieved by the dense retrieval model for a query.
\item \textbf{Goodrate@100}: A relevance experience metric calculating the proportion of items in the top-100 retrieved results by the dense retrieval model that are classified as "Good" (i.e., L4-Excellent, L3-Related) by the relevance model.
\end{itemize}

Next, we provide a concise description of the offline test sets:

\begin{itemize}
\item \textbf{General Retrieval Test Set}: Constructed by annotating items exclusively retrieved by the dense retrieval channel (using the relevance model and manual labeling) and filtering. Contains over 50,000 queries and 2,000,000 query-item pairs. Focus metric: Hitrate@6k.
\item \textbf{Long-Tail Query Test Set}: Constructed by collecting long-tail queries from end-to-end search logs, annotating existing retrieval results (using the relevance model and manual labeling), and filtering. Contains over 10,000 queries and 300,000 query-item pairs. Focus metric: Hitrate@6k.
\item \textbf{Long-Tail Experience Test Set}: Deliberately includes queries from four challenging categories: negation, substitutes, QA, and knowledge-based queries. Contains over 8,000 queries. Focus metric: Goodrate@100.
\end{itemize}

For online evaluation, we conduct online A/B testing and compute the following metrics through manual evaluation:

\begin{itemize}
\item \textbf{GSB (Good/Same/Bad)}: A pairwise comparison metric used in A/B testing. Human evaluators assess whether results from the test method outperform, equal, or underperform those from the baseline method for the same query. GSB+x\% denotes that x\% of test method results are rated superior to the baseline.
\item \textbf{Query Goodrate}: Measures page-level relevance by categorizing entire search result pages into three classes: Good(high relevance), Mid(moderate relevance), or Bad(low relevance). The metric is calculated as the percentage of queries classified as Good or Mid.
\item \textbf{Item Goodrate}: Quantifies item-level relevance by computing the proportion of highly relevant items (e.g., rated 4-Excellent or 3-Related) per search request. The final metric is the average of these proportions across all requests.
\end{itemize}

\subsubsection{Implementation Details}
The model is trained on a cluster of 256 GPUs (each equipped with 96GB memory). The training framework is implemented based on ROLL\cite{wang2025reinforcement}, an open-source distributed RL-training library optimized for large language models. During the Supervised Fine-Tuning (SFT) phase, a per-GPU batch size of 256 is configured with the AdamW optimizer (\(\beta_1\)=0.9, \(\beta_2\)=0.95), a learning rate of 1e-5, and a cosine learning rate scheduler incorporating a warmup ratio of 0.05. The architecture enforces a maximum input length of 256 tokens, with query and item embeddings projected into a 1024-dimensional L2-normalized space. MRL is applied across dimensions [1024, 2048, 3072] to enhance hierarchical representation learning. This phase completes after 1 full training epoch over the dataset.

Transitioning to the Retrieval-GRPO phase, the policy model (3B parameters) interacts with the TaoSR1 as reward model—a 42B Mixture-of-Experts (MoE) architecture activating 3.5B parameters per instance. Maintaining consistency with the SFT phase, the training protocol employs a per-GPU batch size of 256, AdamW optimization (\(\beta_1\)=0.9, \(\beta_2\)=0.95), and a reduced learning rate of 1e-6, coupled with the same cosine scheduler and 256-token context window. The KL divergence penalty coefficient (\(\beta\)=0.2, as defined in Equation \ref{eq:grpo}) regulates policy deviation from reference behavior. Embedding dimensions remain fixed at 1024 to ensure compatibility with upstream components. The GRPO phase iterates over 3 epochs to achieve convergence.


\begin{table}[t]
    \centering
    \caption{Offline Performance Comparison of Training Strategies with Hitrate@6000}
    \begin{tabular}{lccc}
    \toprule
        \multirow{2}{*}{Model} & \multicolumn{2}{c}{Hitrate@6000} \\
        & General & Long-Tail \\ 
        \midrule
        BERT-base & 32.28 & 19.10  \\
        Qwen-3B & 45.08 & 33.67  \\
        Tbstars Post-Pretrain(base) & 46.76 & 32.74 \\ 
        \midrule
        Tbstars SFT Only & 48.00 & 37.52  \\
        Tbstars R-GRPO Only & 46.20 & 37.79  \\
        Tbstars SFT + Ranking loss(Hard Neg) & 48.06 & 38.11  \\
        Tbstars SFT + DPO(Hard Neg) & 48.20 & 39.35  \\
        Tbstars SFT + R-GRPO & 49.35 & 41.07  \\
        
    \bottomrule
    \end{tabular}
    \label{tab:main_result}
\end{table}

\subsubsection{Baselines}
\begin{itemize}
\item BERT-base: A 12-layer 110M BERT trained via the RetroMAE method and contrastive learning.
\item Query-Rewrite: Query rewriting via CSA-QR\cite{feng2025complicated}, followed by retrieval via inverted index.
\item Qwen-3B\cite{yang2025qwen3}: Dense retrieval model trained based on Qwen-3B, modifying to bi-directional attention.
\item Tbstars Post-Pretrain(base): Dense retrieval model trained based on an internal 
closed-source Tbstars with 3B parameters, modifying to bi-directional attention.
\item Tbstars SFT + Ranking loss(Hard Neg): After post-pretrain on Tbstars, performing SFT and pairwise ranking training on Tbstars with hard negative samples.
\item Tbstars SFT + DPO(Hard Neg): After post-pretrain on Tbstars, performing SFT and DPO training with hard negative samples.
\end{itemize}

\subsection{Offline Evaluation}
Offline evaluation is conducted using hitrate@6000 and goodrate@100 metrics. Table \ref{table:main_goodrate} presents the offline experience evaluation results from the view of relevance, and Table \ref{tab:main_result} shows the offline evaluation results from the view of recall.

As shown in Table \ref{table:main_goodrate}, the model with SFT and Retrieval-GRPO achieves optimal performance, outperforming the state-of-the-art baseline Tbstars (SFT only) with an average improvement of 4 pt across four challenging query slices. The most significant enhancement is observed in the alternative querys, showing a 4.62 pt gain. 
It is because the alternatives typically share identical functionality 
with the original items 
but differ in brand and price. From the lexical overlap metric perspective, queries exhibit higher overlap scores with original brands than their alternatives, bringing challenges for models relying primarily on lexical matching. For instance, a search for "alternative PS5 controller" might incorrectly prioritize official Sony products due to brand name overlap, despite the user's explicit request for alternatives. Through Retrieval-GRPO, wrong predictions can be corrected via relevance rewards, which helps the model overcome superficial lexical matches and develop deeper semantic understanding.

As shown in Table \ref{tab:main_result}, the two-stage SFT+Retrieval-GRPO model achieves the best performance, outperforming the current state-of-the-art baseline by 1.35pt and 3.55pt on general queries and long-tail queries, respectively. Additionally, we observe that under the same training framework, Tbstars-3B performs better on general queries while Qwen-3B shows advantages on long-tail queries. We attribute it to Tbstars' incorporation of extensive e-commerce data during pre-training, enabling stronger comprehension of general e-commerce queries. In contrast, understanding long-tail queries relies more on zero-shot capabilities. Qwen retains richer general semantic information, leading to superior performance on long-tail scenarios like question-answering tasks.

\begin{table}[t]
    \centering
    \caption{Ablation Results on Retrieval-GRPO and Relevance Reward}
    \begin{tabular}{lcccc}
    \toprule
        \multirow{2}{*}{Model} & \multicolumn{2}{c}{Hitrate@6000} & \multirow{2}{*}{Quality} \\
        & General & Long-Tail &  \\ 
        \midrule
        Tbstars SFT Only & 48.00 & 37.52 & -  \\
        Tbstars SFT + R-GRPO(Rel Only) & 49.21 & 41.01 & -0.15\% \\
        Tbstars SFT + R-GRPO & 49.35 & 41.07 & +11.64\% \\

    \bottomrule
    \end{tabular}
    \label{tab:abla1}
\end{table}

\begin{table}[]
    \centering
    \caption{Ablation Results on Exclusivity Reward}
    \begin{tabular}{lcc}
    \toprule
        Model & Item Goodrate & Overlap(\%) \\
        \midrule
        Base & - & - \\
        SFT + R-GRPO(w/o Exclusivity) & +9.21pt & 25.7\% \\
        SFT + R-GRPO & +9.14pt  & 17.6\% \\

    \bottomrule
    \end{tabular}
    \label{tab:abla2}
\end{table}

\begin{table}[]
    \centering
    \caption{Ablation Results on Reward Model}
    \resizebox{0.51\textwidth}{!}{
        \begin{tabular}{lcccc}
        \toprule
            \multirow{2}{*}{Model} & \multicolumn{2}{c}{Hitrate@6000} & \multirow{2}{*}{Goodrate@100} \\
            & General & Long-Tail &  \\ 
            \midrule
            SFT & 48.00 & 37.52 & 63.44\% \\
            SFT + R-GRPO(Rel-BERT-4l) & 47.40 & 37.22 & 63.67\% \\
            SFT + R-GRPO(Rel-42BA3.5B) & 49.35 & 41.07 & 65.91\% \\
    
        \bottomrule
        \end{tabular}
    }
    \label{tab:abla3}
\end{table}

\begin{figure*}[t]
    \centering
    \includegraphics[width=0.95\linewidth]{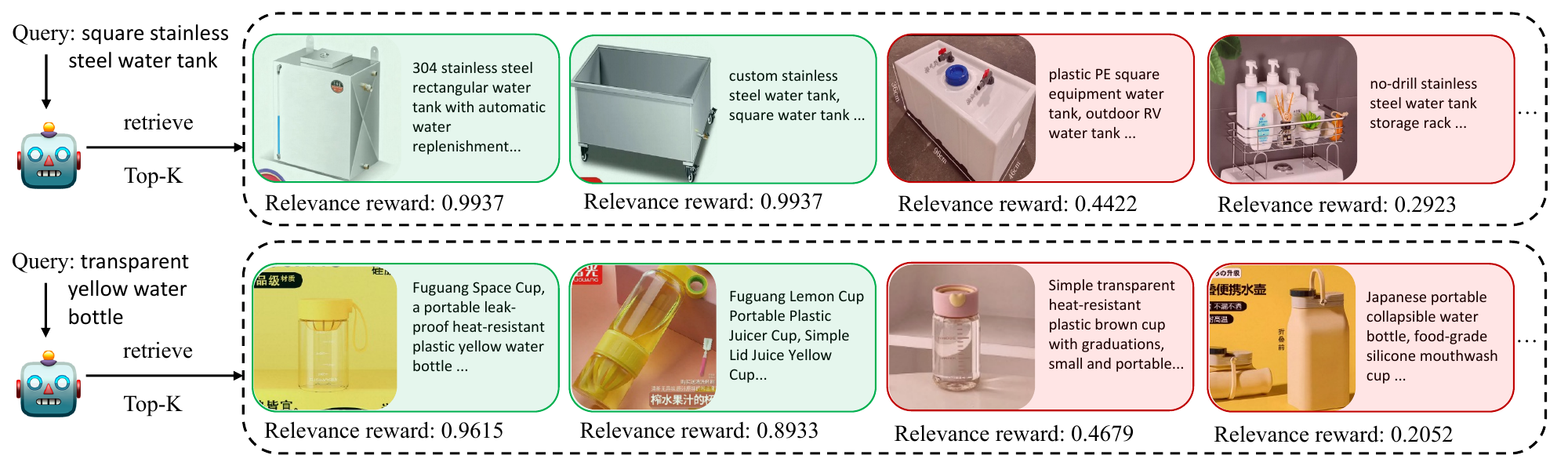}
    \caption{Top-k items in R-GRPO and relevance scores assigned by the relevance model for query-item pairs}
    \label{fig:case1} 
\label{pic:case1}
\end{figure*}

\begin{table*}[t]
    \centering
    \caption{Online Side-by-Side Human Evaluations}
    \label{tab:online_table}
    \begin{tabularx}{0.8\textwidth}{l >{\raggedright\arraybackslash}X c c c}
    \toprule
        Query Type & Case & GSB & Query Goodrate & Item Goodrate  \\ 
        \midrule
        Q\&A & What medicine can make hair black? & +24.30\% & +10.64pt & +8.2pt \\
        Alternative & Alternative PS5 controller & +28.57\% & +10.87pt & +9.1pt \\
        Negative & Short sleeves that don't stick to hair & +41.16\% & +17.07pt & +11.65pt \\
        Knowledge & Paint that is not afraid of car pressure & +25.5\% & +12.65pt & +7.60pt \\
        Overall & - & +29.88\% & +12.81pt & +9.14pt \\
    \bottomrule
    \end{tabularx}
\end{table*}

\begin{figure}[]
    \vspace{-0.1cm}
    \centering
    \includegraphics[width=0.90\linewidth]{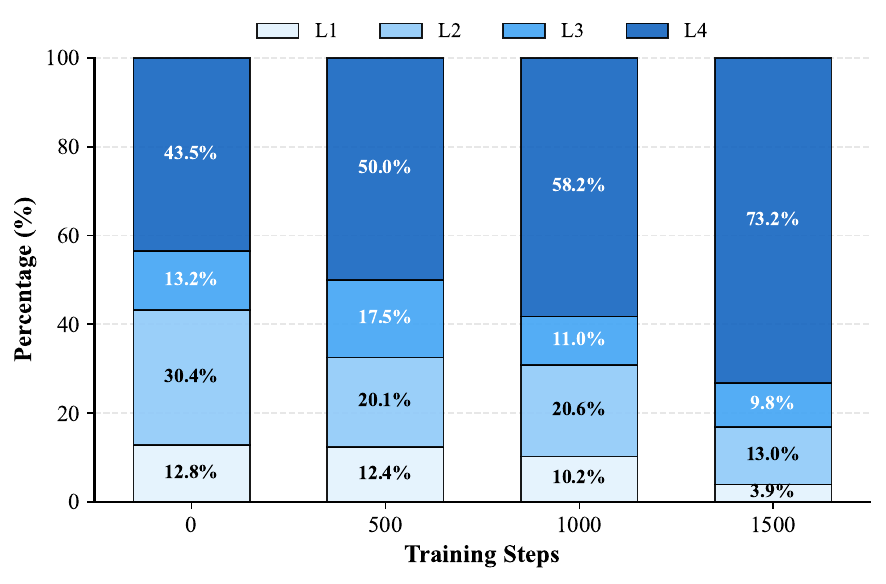}
    \vspace{-0.4cm}
    \caption{Distribution of Relevance Levels during R-GRPO}
    \vspace{-0.6cm}
\label{pic:dist_per}
\end{figure}

\subsection{Ablation Study}
\subsubsection{Effect of R-GRPO}
As shown in Table \ref{tab:main_result}, we conduct ablation studies on the two-stage training framework. The SFT+Retrieval-GRPO approach significantly outperforms the SFT-only baseline. Additionally, replacing the second stage with a ranking paradigm or DPO training with offline hard negative samples both resulted in performance degradation, demonstrating the clear advantages of our proposed Retrieval-GRPO phase. Furthermore, when we removed the first-stage SFT and directly applied reinforcement learning, the performance declined substantially, indicating that while Retrieval-GRPO is effective, the first SFT stage remains essential. The training framework can be viewed as a progressive difficulty training approach: if the model lacks the basic ability to distinguish between simple positive and negative samples, it will struggle even more with challenging samples. 

Figure \ref{pic:dist_per} shows how the distribution of average relevance-score buckets for the top-k items evolves during training. It can be observed that L1 and L2 gradually shift towards L3 and L4 over the course of training.

\begin{figure*}[t]
    \centering
    \includegraphics[width=0.95\linewidth]{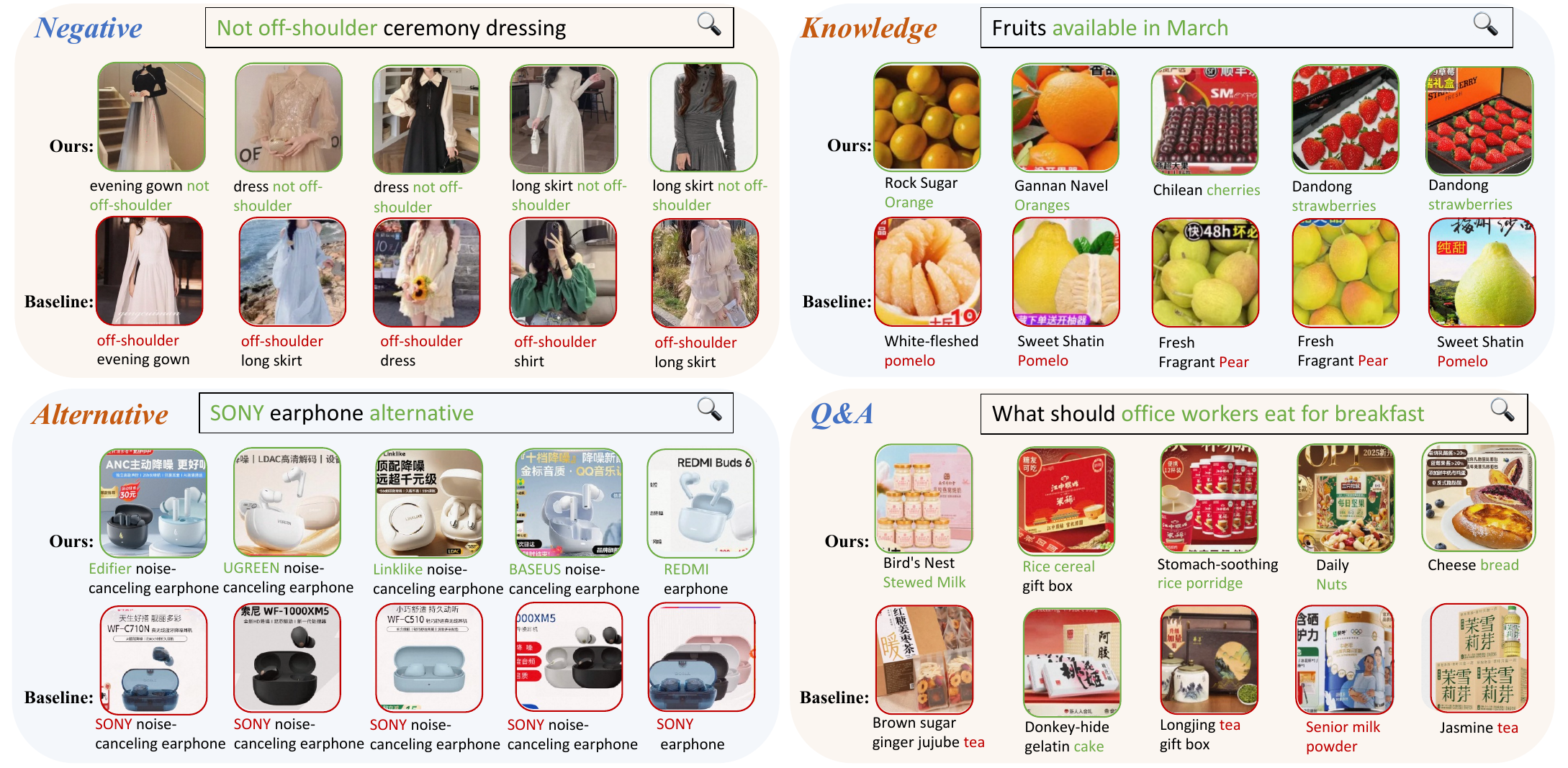}
    \caption{Recall comparison between our model and the online baseline on four long-tail query-slices}
    \vspace{-0.2cm}
    \label{fig:case2}
\end{figure*}

\subsubsection{Effect of Multi-Objective Reward}
To demonstrate the effectiveness of multi-objective rewards, we conducted multi-dimensional ablation experiments to verify the efficacy of item quality reward and exclusivity reward beyond the relevance dimension. As shown in Table \ref{tab:abla1}, compared with the SFT-only model, when training with only relevance score reward in Retrieval-GRPO, the average quality score of retrieved products remained unchanged, but the Hitrate@6000 metric showed significant improvement. Using combined relevance+quality+exclusivity rewards for multi-objective training achieved notable improvements in Hitrate@6000 along with an 11.64\% increase in average item quality score. Table \ref{tab:abla2} reveals the overlap ratio between items retrieved by dense retrieval and inverted index pipelines. Without exclusivity reward training, the overlap ratio with the inverted index way reached 25.7\%, which decreased to 17.6\% after incorporating exclusivity reward training. These experiments collectively demonstrate the effectiveness of multi-objective reward reinforcement learning training.

\subsubsection{Effect of Reward Model}
To validate the role of relevance model-generated reward scores in Retrieval-GRPO, we conducted ablation experiments on the Reward Model. In our training setup, we utilized TaoSR1 (a relevance model developed from the Tbstars-42BA3.5B via reinforcement learning) as the default reward model. In the experiment, we replaced it with a 4-layer BERT-based relevance model. As shown in Table \ref{tab:abla3}, using the 4-layer relevance model as the reward model yielded almost no performance gains, while training with TaoSR1 significantly outperformed the baseline. It demonstrates that precise relevance reward calibration is pivotal for Retrieval-GRPO. Inaccurate reward assignments may degrade model performance.

\subsection{Case Study}
\subsubsection{Examples in R-GRPO}
Through R-GRPO, the model can automatically mine hard negative samples during training and perform debiased learning based on the relevance scores. Figure \ref{fig:case1} illustrates the top-k items selected by the dense retrieval model during R-GRPO training, along with the relevance scores assigned by the relevance model to these query-item pairs. As demonstrated in the figure, for the user search query "square stainless steel water tank", among the top-k items, two stainless steel square sinks get high relevance scores (0.99 and 0.99), while a plastic product (0.44) and a sink storage rack (0.29) obtain significantly lower relevance scores.

\subsubsection{Case Studies on long-tail Query}
Figure \ref{eq:score2} presents a comparison between our model and the baseline currently deployed online across four types of long-tail queries. As can be seen, compared with the baseline, with R-GRPO our model can better distinguish positive samples from hard negative samples, which benefits from real-time feedback learning during the training process.
\vspace{-0.2cm}

\subsection{Online Evaluation}
We deploy the model into the real-world pipeline and validated its effectiveness in the online search engine through A/B testing. A side-by-side evaluation was conducted between two experimental buckets, where the top 10 items for 2000 actual 
online queries across four challenging query slices are manually annotated using three human evaluation metrics introduced in subsection \ref{exp.setup}.

The specific results are shown in Table \ref{tab:online_table}. We conducted fine-grained analysis on four selected challenging query slices. The most significant improvement was observed in negative-class queries, as negative semantics and literal overlap metrics typically exhibit opposing trends, posing additional challenges for models that rely on literal matching rather than deep semantic understanding for traditional dense retrieval model. Through Retrieval-GRPO, the model leverages the semantic generalization capabilities of relevance models based 
on LLM and employs 
real-time relevance feedback on retrieved top-items via reinforcement learning during training, enabling timely correction of wrong predictions.

These results demonstrate that through the multi-objective reinforcement learning retrieval framework of Retrieval-GRPO training, the model can fully unleash the potential of large language models, significantly improving user search experience for challenging queries requiring deep understanding of user intent.

The model has now been fully integrated into Taobao's search pipeline.

\section{Conclusion}
This paper presents Retrieval-GRPO, a multi-objective reinforcement learning framework designed to address the limitations of conventional dense retrieval paradigms in e-commerce search systems. By leveraging dynamic candidate retrieval and multi-objective reward fusion in RL paradigm, our framework eliminates the dependency on labor-intensive hard negative sample mining while mitigating the seesaw effect inherent in multi-objective optimization. The proposed method integrates real-time relevance feedback from an LLM-based reward model, 
item quality metrics, and exclusivity
constraints, and dynamically optimizes semantic representations through reinforcement learning.
Extensive experiments demonstrate that Retrieval-GRPO significantly enhances semantic generalization capabilities, particularly for complex long-tail queries, by enabling real-time error correction and preference alignment during training. The framework’s elimination of static hard negative mining pipelines not only accelerates iteration cycles but also ensures continuous learning from feedback on candidates sampled by the latest gradient-updating model. Furthermore, the unified multi-objective reward mechanism provides a more flexible and effective optimization strategy compared to traditional multi-loss joint training.
Deployed on China’s largest e-commerce platform, Retrieval-GRPO validates its practical efficacy in improving search relevance and transaction conversion rates.

\clearpage
\bibliographystyle{ACM-Reference-Format}
\bibliography{paper}

\clearpage
\appendix
\begin{figure*}[]
    \centering
    \includegraphics[width=\linewidth]{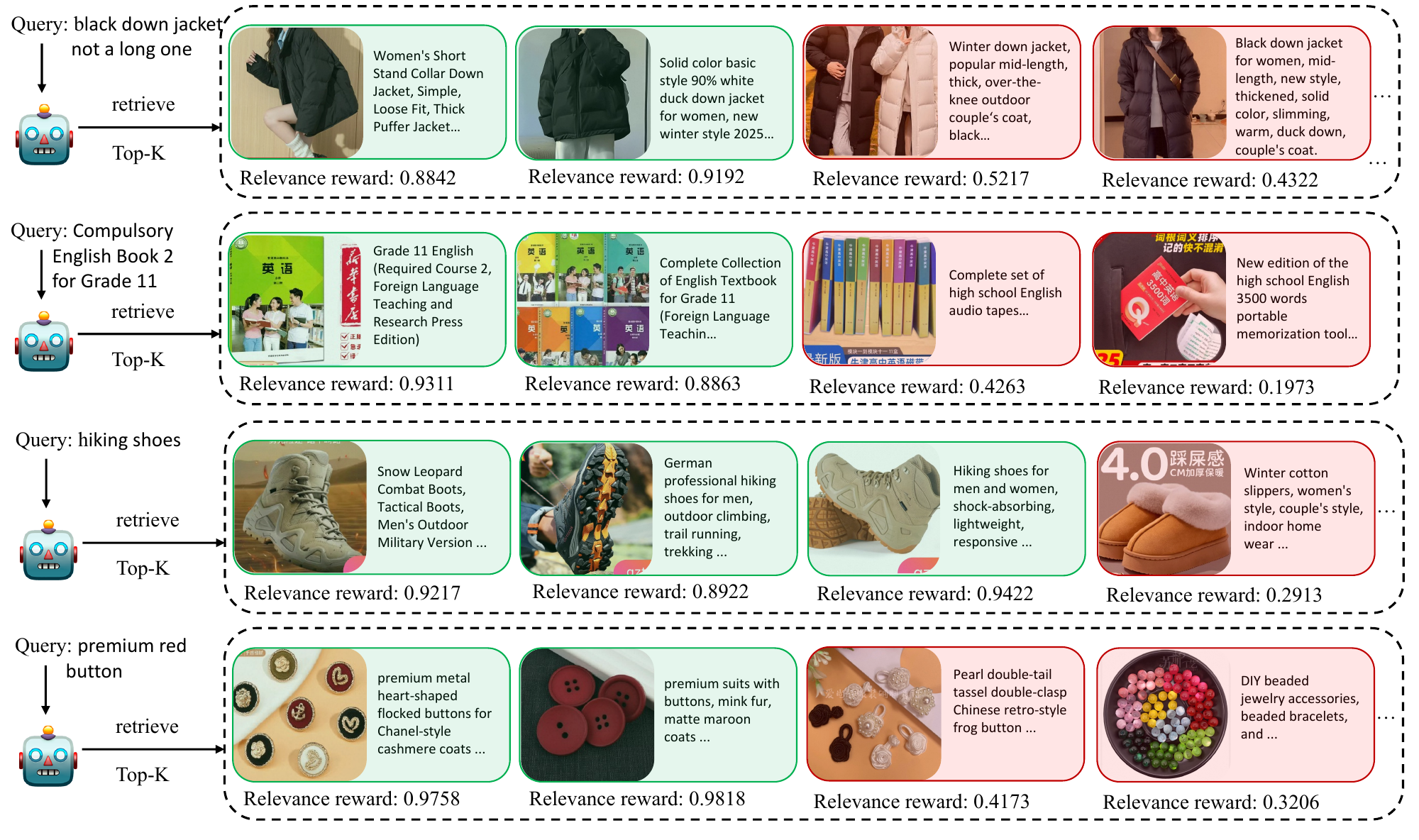}
    \caption{Top-k items in R-GRPO and relevance scores assigned by the relevance model for query-item pairs}
    \label{fig:case1_appendix}
\end{figure*}

\begin{figure*}[]
    \centering
    \includegraphics[width=\linewidth]{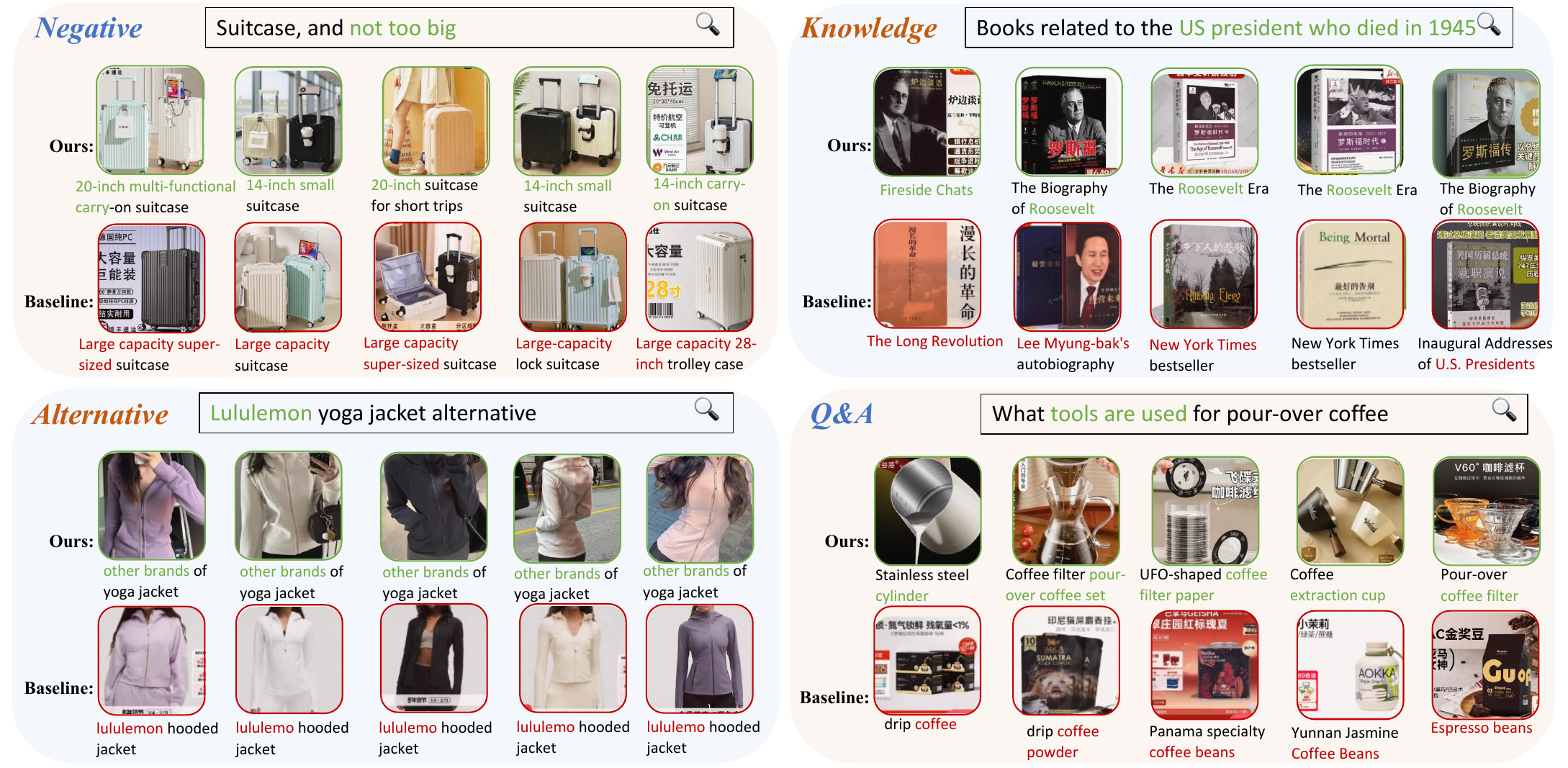}
    \caption{Recall comparison between our model and the online baseline on four long-tail query-slices}
    \label{fig:case2_appendix}
\end{figure*}

\end{document}